\documentclass[12pt,preprint]{aastex}

 \slugcomment{submitted to Astrophys. J., January 31, 2001}
 \shorttitle{Measurement of the \ion{C}{4} rate coefficient}
 \shortauthors{S.\ Schippers et al.}

\begin{document}

 \title{Storage ring measurement of the \ion{C}{4} recombination rate coefficient}

 \author{S. Schippers and A. M{\"u}ller}

 \affil{Institut f{\"u}r Kernphysik, Strahlenzentrum der Justus-Liebig-Universit{\"a}t, 35392 Giessen, Germany}

 \author{G. Gwinner, J. Linkemann, A. A. Saghiri, and A. Wolf}

 \affil{Max-Planck-Institut f{\"u}r Kernphysik, 69117 Heidelberg, Germany
                 and Physikalisches Institut der Universit\"{a}t Heidelberg, 69120 Heidelberg, Germany}

\begin{abstract}
The low energy \ion{C}{4} dielectronic recombination (DR) rate coefficient associated with
2s$\to$2p $\Delta n=0$ excitations of this lithiumlike ion has been measured with high
energy-resolution at the heavy-ion storage-ring {\sc tsr} of the Max-Planck-Institut f{\"u}r
Kernphysik in Heidelberg, Germany. The experimental procedure and especially the
experimental detection probabilities for the high Rydberg states produced by the
recombination of this ion are discussed in detail. From the experimental data a Maxwellian
plasma rate coefficient is derived with $\pm15\%$ systematic uncertainty and parameterized
for ready use in plasma modeling codes. Our experimental result especially benchmarks the
plasma rate coefficient below $10^4$ K where DR occurs predominantly via
\ion{C}{3}(1s$^2$2p4l) intermediate states and where existing theories differ by orders of
magnitude. Furthermore, we find that the total dielectronic and radiative \ion{C}{4}
recombination can be represented by the incoherent sum of our DR rate coefficient and the RR
rate coefficient of Pequignot et al. (1991, Astron.\ Astrophys., 251, 680).
\end{abstract}

\keywords{atomic data --- atomic processes --- recombination --- X-rays: general}

\section{Introduction}

Carbon is one of the cosmically most abundant elements. Consequently, line emission from
carbon ions is observed from a wide range of cosmical objects. In low density, photoionized
and electron-ionized cosmic plasmas the dominant mechanisms for recombination are radiative
recombination (RR) and dielectronic recombination (DR). For \ion{C}{4} these processes can
be expressed as
\begin{equation}
{\rm C}^{3+}(1s^22s) + e^- \to {\rm C}^{2+}(1s^22snl) +
h\nu\label{eq:RR}
\end{equation}
and
\begin{equation}
 {\rm C}^{3+}(1s^22s) + e^- \to {\rm C}^{2+}(1s^22pnl) \to
 \left\{\begin{array}{l} {\rm C}^{2+}(1s^22snl)+ h\nu \\
                         {\rm C}^{2+}(1s^22pn'l')+ h\nu',
        \end{array}\right.\label{eq:DR}
\end{equation}
respectively. In low energy \ion{C}{4} DR the $1s^22s$ ion is excited by a $2s\to 2p$
$\Delta n=0$ transition (excitation energy $\sim 8$~eV) only within the $n=2$ shell. At
higher energies above 37.55~eV, the $2s$ electron can be excited to $3l$ substates ($\Delta
n = 1$). The lowest-energy resonances of the associated DR series with configurations $1s^2
3l3l'$ of \ion{C}{3} are at energies above about 12~eV. The strengths of these resonances
can be expected to be very much smaller than those associated with $2s\to2p$ core
transitions. The K-shell cannot be excited with energies smaller than 240~eV. In this paper
we exclusively deal with $\Delta n=0$ DR.

The calculation of DR rate coefficients is a challenging task since an infinite number of
states is involved in this process. Moreover, relativistic and many-body effects have to be
accounted for in high orders even in the case of DR of a light ion such as \ion{C}{4}
\citep{Mannervik1998}. Approximations and computational simplifications are needed in order
to make DR calculations tractable. It turns out that different calculations yield rate
coefficients differing by up to orders of magnitude. In this situation benchmarking
experiments are vitally needed in order to guide the development of the theoretical methods
and to provide reliable DR rate coefficients for plasma modelers. For \ion{C}{4} the
available theoretical rate coefficients have recently been critically compared by
\citet{Savin2000c}.

In the past decade electron coolers at heavy-ion storage rings have developed into the most
successful experimental tool for electron-ion recombination studies
\citep{Schuch1993,Mueller1997c}. Currently, corresponding research programmes are carried
out at the heavy ion storage rings {\sc esr} of the Gesellschaft f\"{u}r Schwerionenforschung
(GSI) in Darmstadt \citep{Brandau1997}, {\sc tsr} of the Max-Planck-Institut f\"{u}r Kernphysik
in Heidelberg \citep{Mueller1997b,Mueller1998,Wolf2000} and {\sc cryring} of the
Manne-Siegbahn-Laboratory in Stockholm \citep{Schuch1997,Schuch1998}. Extensive
bibliographic compilations on DR measurements at storage rings have been published e.g.\ by
\citet{Mueller1995} and \citet{Schippers1999b}. The basic approach for deriving plasma rate
coefficients from storage-ring measurements has been summarized by \citet{Mueller1999c}.
Recent experimental work on plasma rate coefficients for astrophysical and other plasma
applications has been published by \citet{Savin1997,Savin1999} on DR of \ion{Fe}{18} and
\ion{Fe}{19}, by \citet{Schippers1998} on DR of \ion{Ti}{5} and by \citet{Schippers2000b} on
DR of \ion{Ni}{26}.

In this paper we present the measured \ion{C}{4} recombination rate coefficient. It is
organized as follows. In Sec.~\ref{sec:exp} the experimental procedure is described.
Particular attention is given to the field ionization of high-$n$ Rydberg states in motional
electric fields that are unavoidable with the experimental arrangement at an ion
storage-ring. In Sec.~\ref{sec:res} the experimental result is presented and the impact of
field ionization on the measured DR resonance strength is discussed. A theoretical estimate
of the unmeasured DR rate is presented, and the \ion{C}{4} plasma DR rate coefficient is
derived. In Sec.~\ref{sec:discDR} it is compared to existing theoretical results. Finally,
in Sec.~\ref{sec:total} we derive the total DR+RR \ion{C}{4} rate coefficient and compare it
to a unified calculation of \citet{Nahar1997}. A model calculation of $nl$-selective
detection probabilities for recombined ions in high-$n$ Rydberg states, that takes the field
ionization properties of our experimental apparatus into detailed account, is presented in
appendix \ref{sec:dp}.

\section{Experiment}\label{sec:exp}

The \ion{C}{4} recombination measurements have been performed at the heavy-ion storage ring
{\sc tsr} \citep{Jaeschke1989} of the Max-Planck-Institut f\"ur Kernphysik in Heidelberg,
basically following the procedure of earlier measurements \citep{Kilgus1992,Lampert1996}. A
beam of $^{12}$C$^{3+}$ ions at an ion energy $E_{\rm i} \approx 1.5$~MeV/u was supplied by
the MPI accelerator facility and injected into the {\sc tsr}. In the {\sc tsr} electron
cooler, situated in one of the straight sections of the storage ring (cf.\
Fig.~\ref{fig:exp}), the circulating C$^{3+}$ ions were merged with a beam of electrons
moving collinearly with the ions at roughly identical velocity. At matched beam velocities
the electron beam had an energy $E_{\rm c} = (m_{\rm e}/m_{\rm i})E_{\rm i} \approx 840$~eV.
It was merged with the ion beam over a length of 1.5~m and was guided by a longitudinal
magnetic field of 42~mT over its entire path. The electron density was $5-7\times
10^6$~cm$^{-3}$ depending on the laboratory electron energy $E_{\rm e}$. The electron beam's
diameter was 3 cm. Electron cooling, i.e., the ion interaction with the overlapping electron
beam, on a time scale of $\approx1$~s compresses the circulating ion beam to a diameter of
1--2 mm; simultaneously the longitudinal velocity of the freely coasting ions adjusts itself
to the average electron velocity. By repeated injections, while keeping the cooled beam
stored, C$^{3+}$ ions were accumulated \citep{Grieser1991} on a time scale of $\approx60$~s
up to electrical currents of typically 30~$\mu$A, corresponding to $\approx 8\times10^7$
stored ions.

After ion accumulation, recombination rates of C$^{3+}$ ions were
measured by counting recombined C$^{2+}$ ions on a
multi-channelplate detector located behind the first dipole magnet
(bending radius 115~cm) downstream of the electron cooler (see
Fig.~\ref{fig:exp}). The dipole magnet keeps the circulating
C$^{3+}$ ion beam on a closed orbit while it deflects the recombined
C$^{2+}$ ions less strongly so that they hit the detector. Absolute rate coefficients for the
recombination of C$^{3+}$ ions with electrons in the collinear
overlap region were measured as a function of the average relative
energy
\begin{equation}
\hat{E} \approx \left(\sqrt{E_{\rm e}}-\sqrt{E_{\rm
c}}\,\right)^2\label{eq:erel}
\end{equation}
between electron and ion beam. The relativistically correct expression \citep[see
e.g.][]{Schippers2000b} has been used in the data analysis. The laboratory electron energy
$E_{\rm e}$ was deduced from the cathode voltage and electron current applying a correction
for the electron space-charge \citep{Kilgus1992}.

The electron motion in the transverse degrees of freedom, confined by the magnetic field, is
largely decoupled from that in longitudinal direction. In a co-moving reference frame, the
electron velocity spread is characterized by the longitudinal and transverse temperatures
$T_{||}$ and $T_{\perp}$ with $k_{\rm B}T_{\perp}\approx 10$ meV and $k_{\rm B}T_{||}\approx
0.15$ meV $\ll k_{\rm B}T_{\perp}$ ($k_{\rm B}$ denotes the Boltzmann constant). Both
temperatures are considerably lower than the cathode temperature, the low longitudinal
temperature resulting from the electron acceleration and the low transverse temperature
resulting from adiabatic magnetic expansion of the electron beam \citep{Pastuszka1996}. The
electron velocity distribution is represented by
\begin{equation}
f(\hat{v},\vec{v}) =
  \sqrt{\frac{m_{\rm e}}{2\pi k_{\rm B}T_{\parallel}}}
    \exp\left[{-\frac{m_{\rm e}(v_{\|}-\hat{v})^2}{2k_{\rm B}T_{\parallel}}}\right] \frac{m_{\rm e}}{2\pi k_{\rm B}T_{\perp}}
           \exp\left({-\frac{m_{\rm e} v_{\perp}^2}{2k_{\rm B}T_{\perp}}}\right).
\label{eq:fele}
\end{equation}
The experimental energy resolution corresponds to the width of this distribution and amounts
to \citep{Mueller1999c}
\begin{equation}
\Delta \hat{E}(\mbox{\rm FWHM}) = \left[\left(\ln(2)\,k_{\rm B}T_\bot\right)^2 + 16\ln(2)\,
k_{\rm B}T_\|\,\hat{E}\right]^{1/2}.\label{eq:deltae}
\end{equation}

As has been demonstrated by \citet{Bartsch1999a,Bartsch2000} and \citet{Schippers2000b}, DR
rate coefficients can considerably be influenced by external {\it electric} and additional
crossed {\it magnetic} fields in the TSR electron-cooler. Therefore, care has been taken to
minimize such fields in the interaction region. Additional to the magnetic field of 42~mT
only small electric stray fields are present in the interaction region. Components of the
magnetic field transverse to the ion velocity produce motional electric fields in the ion
rest frame amounting to 2~V/cm for the estimated maximum field angles of about 0.3~mrad.
Small stray fields are also expected from the electron space charge \citep{Kilgus1992}. By
monitoring the velocity of the stored, electron-cooled ion beam, the ions are centered
within the electron beam to about $\pm2$~mm. For aligned electron and ion beams the electric
space-charge field remains below 1~V/cm over the ion beam cross section of $\leq2$~mm
diameter. Altogether, electric stray fields below 3~V/cm are estimated. Considering the
findings by \citet{Bartsch1999a,Bartsch2000} and \citet{Schippers2000b}, we conclude that a
reasonably 'field-free' DR measurement can be carried out under these conditions.

Before their detection the recombined C$^{2+}$ ions have to travel through the toroid magnet
that guides the electron beam out of overlap, through correction dipoles, quadrupoles, and
through the deflection dipole that separates them from the stored C$^{3+}$ beam
(Fig.~\ref{fig:exp}). All these components cause transverse magnetic fields considerably
higher than those in the interaction region, leading to motional electric fields that can
field-ionize recombined ions in highly excited states. Since a sizable fraction of
recombined ions is expected to be formed in highly excited states, the efficiency of their
detection needs particular consideration. Recombined ions that reach a zone of large
motional electric fields in sufficiently excited states will be field ionized and hence
cannot be detected; however, ions formed in highly excited states that are able to decay
radiatively to lower levels before reaching the critical field region will still be
detected. At the given measuring conditions, the 1.5-m nominal length of the interaction
region corresponds to a flight time of 87~ns. The motional electric field rises to $\sim$4
kV/cm in the toroid after a flight time of $55\pm44$~ns, depending on the location along the
interaction path where the recombination takes place. Two correction magnets with peak
fields of $\sim$6~kV/cm and $\sim 12$~kV/cm, are reached after $103\pm44$~ns and $133\pm
44$~ns, respectively, and the deflection dipole with a peak field of $\sim$106 kV/cm after
$275\pm44$~ns. The critical quantum number for field ionization of an ion in an electric
field $F$ is estimated from \citep{Gallagher1994}
\begin{equation}
n_F = \sqrt[4]{\frac{q^3}{9F}} \label{eq:nf}\end{equation} where $q$ is the charge of the
ion core and $F$ is measured in atomic units (1 au = $5.142\cdot10^9$~V/cm). For
C$^{2+}(nl)$ the core charge state is $q=3$ and thus the critical quantum number is $n_F=19$
for $F=106$~kV/cm. On the other hand, the radiative lifetime of the $19p$ level in C$^{2+}$
is 16~ns (hydrogenic value) and hence considerably shorter than the 275~ns flight time from
the center of the cooler to the charge analysing dipole magnet. The detection probability
for recombination events leading to high Rydberg levels $n\gtrsim20$ will therefore depend
on the competition between the radiative decay and the flight time to regions where motional
electric fields are present. Reciprocally, the experimental setup offers a detection
probability near unity for all recombination events leading into states with principal
quantum numbers at least up to about 20.

With a cooled and accumulated beam of C$^{3+}$ ions of typically 30~$\mu$A freely
circulating in the ring, measurements of the recombination rate were performed by stepping
the acceleration voltage in the electron cooler to values different from those at cooling
for short time intervals and recording the detector count rate after the acceleration
voltage had settled to a stable value. In sequence, the acceleration voltage was set to the
value required to obtain the desired relative energy where the recombination rate
coefficient was to be measured, and then to a value producing a fixed, high relative energy
where the count rate of C$^{2+}$ ions was dominated by capture in the residual gas and where
a background rate was measured. The corresponding time intervals will be called `signal' and
`reference' windows, respectively. The settling time was 0.5 ms and the measuring time at a
given acceleration voltage was typically 20 ms. After the signal and reference intervals the
acceleration voltage was stepped back to the cooling value and kept there for 30 ms to
ensure a fixed ion energy $E_{\rm i}$ and a well-cooled ion beam for each measurement.
Measurement cycles were repeated scanning the acceleration voltage level applied during the
`signal' window. A scan comprising 420 data points took $\sim30$~s and during that time the
stored C$^{3+}$ ion current decreased only slightly; the ion current was restored by
`stacking' a few new injections and several such scans were repeated until the desired
integral ion counts were collected for the recombination spectrum. Our experimental range of
relative energies (0--10.5~eV) was covered by three overlapping scans.

The raw data consist of time correlated `signal' and `reference' count rates $R^{\rm
s}(E_{\rm e})$ and $R^{\rm r}(E_{\rm e})$ for each value of the laboratory electron energy
$E_{\rm e}$ reached during the scans. Both rates include the background from electron
capture of C$^{3+}$ ions in the residual gas, and since they are measured in time windows
only milliseconds apart the background is expected to cancel in the difference $R^{\rm
s}(E_{\rm e})-R^{\rm r}(E_{\rm e})$ even in the case of variations of the residual gas
pressure on a time scale as fast as seconds. After conversion to the scale of relative
energies, $\hat{E}$, the energy dependent recombination rate coefficient $\alpha(\hat{E})$
is obtained through
\begin{equation}
\alpha(\hat{E}) = \frac{[R^{\rm s}(\hat{E})-R^{\rm r}(\hat{E})]\,
\gamma^2}{n_{\rm e} N_{\rm i} (L/C) \eta} + \alpha(\hat{E}^{\rm
r})\,\frac{n_{\rm e}^{\rm r}}{n_{\rm e}}, \label{eq:alpha}
\end{equation}
where $n_{\rm e}$ and $N_{\rm i}$ denote the electron density and the number of stored ions,
respectively, $L=1.5$~m is the nominal length of the interaction region, and $C=55.4$~m is
the circumference of the storage-ring. At the relatively high experimental ion energy the
efficiency of the channelplate recombination detector can be assumed to be $\eta = 1$. The
factor $\gamma$ accounts for the relativistic transformation between laboratory and
center-of-mass frames. The second term in Eq.~(\ref{eq:alpha}) is a small correction that
re-adds the electron-ion recombination rate at the reference point. With a suitably chosen
reference point --- $\hat{E}^{\rm r}=10.5$~eV in the present case --- this rate is due to RR
only. Here it is calculated using a modified semi-classical formula for the RR cross section
[see Eq.~(\ref{eq:sigmaBethe}) below] to be $\alpha(\hat{E}^{\rm r})\approx
3.5\times10^{-13}$~cm$^3$s$^{-1}$. Since this is only a small correction to the experimental
data at lower energies the insertion of a rough theory value is justified.

In the toroidal magnetic sections of the electron cooler the
electron beam was guided in and out of the interaction region on
two 45$^\circ$-bends with a radius of 80~cm (cf.\
Fig.~\ref{fig:exp}). In each of these sections, the ion beam
continues to interact with the electron beam at increasing angles
between the two beams and thus shifted relative energies over
regions of $\approx20$~cm length each. From the known geometry,
corrections for these merging regions are included in the data
evaluation \citep{Lampert1996}.

The experimental collision energy ($\hat{E}$) scale is known under the conditions of the
present measurement within a systematic relative error of about 1\%; even higher energy
accuracies were reached after a precise adjustment of the energy scale to known
spectroscopic Rydberg series limits (see below). The systematic uncertainty in the absolute
recombination rate coefficient is estimated to be $\pm15\%$, where the dominant errors are
due to the ion and electron current determination and the detection efficiency. The
influence of the inaccurate knowledge of the effective overlap length $L$ is strongly
reduced by the toroid correction \citep{Lampert1996}. The statistical uncertainty of the
results presented below amounts to less than $1\%$ in the rate coefficient maximum.

\section{Results}\label{sec:res}

\begin{figure}
\plotone{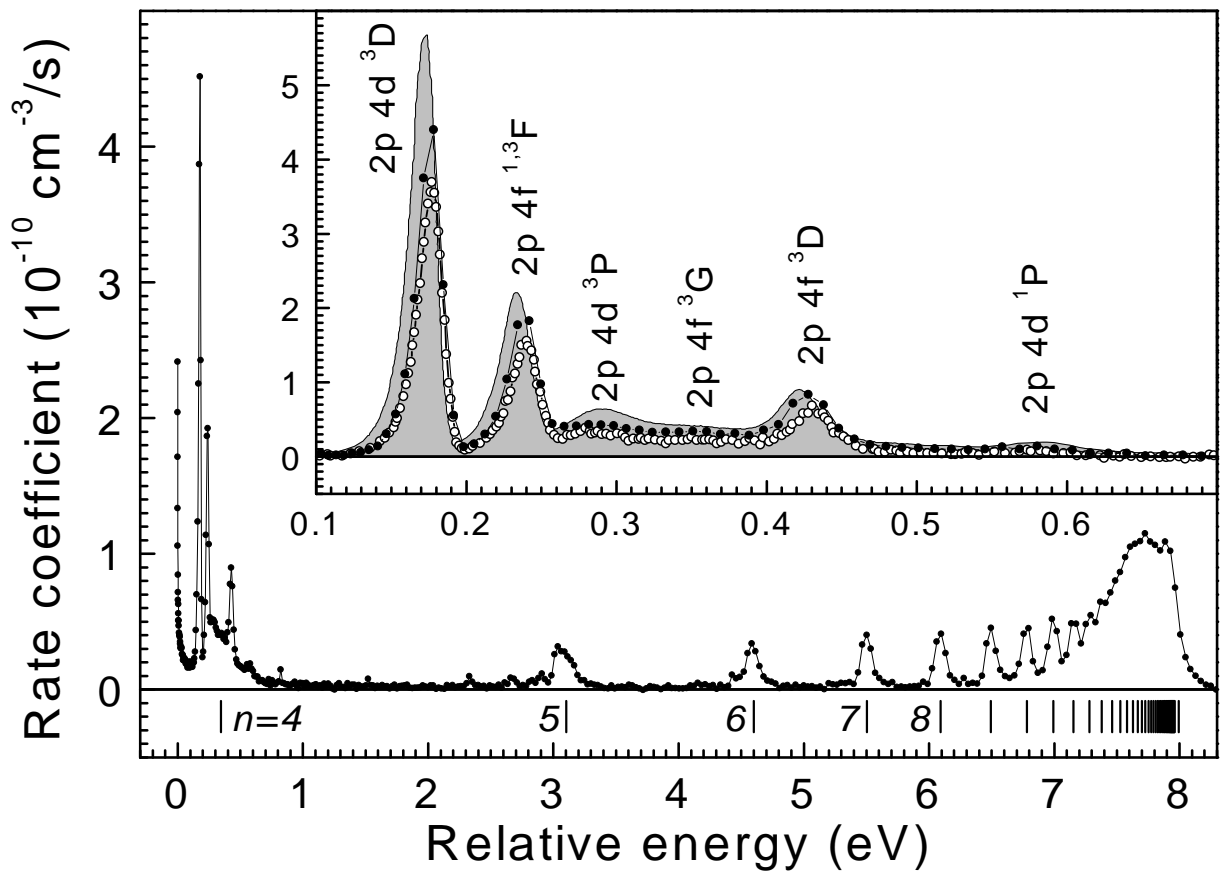} \caption{\label{fig:overview}Experimental \ion{C}{4} rate coefficient
(full circles). The vertical dashes give the $2pnl$ resonance positions according to the
Rydberg formula. The inset shows the background-subtracted (see text) \ion{C}{4} DR spectrum
in the region of the $2p4l$ resonances, the background-subtracted {\sc cryring} data (open
circles), and the theoretical results (shaded curve) of \protect\citet{Mannervik1998}. Their
peak designations are given for the most prominent DR resonances.}
\end{figure}

Our experimental \ion{C}{4} rate coefficient comprising RR and DR contributions is displayed
in Fig.~\ref{fig:overview}. The RR peak at $\hat{E}=0$~eV and individual members of the
$2pnl$ Rydberg series of DR resonances are resolved for $4\leq n\leq 12$. As the
$2p_{1/2}-2p_{3/2}$ splitting in \ion{C}{4} amounts to only 0.013~eV \citep{Edlen1983}, it
is not possible to observe two separate Rydberg series with the given experimental
resolution, in contrast to measurements on heavier lithiumlike ions
\citep{Kilgus1992,Schippers2000b,Bartsch2000}. The series limit corresponding to the
$2s\to2p$ excitation energy was determined by extrapolating the observed Rydberg resonance
positions from $n=5-12$ to infinite $n$. In Fig.~\ref{fig:overview} and throughout the paper
we have adjusted the collision energy ($\hat{E}$) scale by multiplying it with a constant
factor differing from unity by 0.65\%, which brings the extrapolated series limit in
agreement with the spectroscopic value of 8.005~eV \citep{Edlen1983}.

In order to obtain the \ion{C}{4} DR rate coefficient, the RR contribution has been removed
from the spectrum of Fig.~\ref{fig:overview} by subtracting an empirical function
$\alpha_{\rm RR}(\hat{E}) = a_0 + a_1 \hat{E} + a_2/(1+a_3 \hat{E} + a_4 \hat{E}^2)$ from
the experimental rate coefficient with the coefficients $a_i$ determined by fitting
$\alpha_{\rm RR}(\hat{E})$ to those parts of the spectrum which do not exhibit DR
resonances. The corresponding energy intervals which have been used in the fit were
0.021--0.1~eV, 1.0--2.0~eV and 8.5--10.5~eV. The parameters obtained from the fit are $a_0 =
1.556\times10^{-12}$~cm$^3$s$^{-1}$, $a_1 = -1.929\times10^{-13}$~cm$^3$s$^{-1}$eV$^{-1}$,
$a_2 = 4.726\times10^{-11}$~cm$^3$s$^{-1}$, $a_3 = 20.89$~eV$^{-1}$ and $a_4 =
-3.665\times10^{-5}$~eV$^{-2}$. With these parameters $\alpha_{\rm RR}(\hat{E})$ is defined
in the range 0.0021~eV$ \leq \hat{E} \leq $10.5~eV.

In the inset of Fig.~\ref{fig:overview} we compare the part of the DR spectrum that
comprises the $2p4l$ resonances with the measurement of \citet{Mannervik1998} carried out at
the heavy-ion storage ring {\sc cryring} in Stockholm. The published \ion{C}{4} {\sc
cryring} data only extend to $\hat{E}= 0.7$~eV. For the comparison we have removed the RR
background from the present and from the {\sc cryring} measurement by using the procedure
described above. The comparison shows that the two measurements have a similar energy
resolution. The overall shape of the spectrum and the peak positions are almost the same.
However, when integrating over the 0.1--0.7~eV energy range we obtain for the $2p4l$
manifold the resonance strength $1.9\times10^{-11}$~eVcm$^3$s$^{-1}$ and
$2.5\times10^{-11}$~eVcm$^3$s$^{-1}$ from the {\sc cryring} and from the present {\sc tsr}
measurements, respectively. This difference is just within the 30$\%$ summed uncertainty
(15$\%$ for each experiment individually, assuming a similar systematic uncertainty for the
{\sc cryring} measurement as for the present one) for the absolute value of the measured
rate coefficient. Our value is somewhat closer to the theory value
$3.3\times10^{-11}$~eVcm$^3$s$^{-1}$ of \citet{Mannervik1998}.

For the derivation of a meaningful plasma rate coefficient from our experimental data we
have to estimate how much DR strength is not measured due to the cut-off of high Rydberg
states in the electric fields on the recombined ions' path from the cooler to the detector.
To this end we have carried out \ion{C}{4} DR calculations using the atomic structure code
{\sc autostructure} \citep{Badnell1986}. In order to be able to compare the calculated DR
cross section $\sigma$ with our measured rate coefficient, we have performed the convolution
\begin{equation}
\alpha(\hat{E})=\sum_{nl} \Upsilon_{nl}\int \,
\sigma_{nl}\bigl(E(v)\bigr)\,v\,f(\hat{v},\vec{v})\,d^3v,
\label{eq:faltung}
\end{equation}
where we have used the electron velocity distribution of Eq.~(\ref{eq:fele}) with $k_{\rm
B}T_\| = 0.15$~meV and $k_{\rm B}T_\bot = 10$ meV. In Eq.~(\ref{eq:faltung}) $\Upsilon_{nl}$
is the $nl$-specific detection probability of a recombined ion with the outer electron being
in a Rydberg state characterized by the quantum numbers $n$ and $l$. Correspondingly,
$\sigma_{nl}$ denotes the cross section for DR via this Rydberg state. When comparing the
resonance strengths of the individually resolved $2pnl$ DR resonances with $5\leq n\leq10$
we find that the calculation yields values that are on the average a factor 1.25 higher than
the experimental ones. Therefore we have scaled down the calculation by a factor 0.8.
Additionally, the theoretical energy scale has been adjusted by shifting it by 0.06~eV
towards higher energies such that the spectroscopic value for the series limit is
reproduced.

The comparison between our scaled {\sc autostructure} calculation and our experimental
result is shown in Fig.~\ref{fig:theory}. The different curves correspond to different
assumptions for the detection probabilities of high Rydberg states. The full line has been
obtained by setting $\Upsilon_{nl}=1$ for all $nl$ up to $n=1000$ which is the maximum $n$
used in the calculation. By inclusion of such high-$n$ states we have made sure that the
calculation has converged, i.e., calculations up to $n=900$ and $n=1000$ yield rate
coefficients which are undistinguishable from one another. It is obvious that the larger
fraction of the resonance strength due to DR via high Rydberg states has not been measured.
It should be noted that this effect becomes much less significant for DR on more highly
charged ions where comparatively less DR strength is accumulated in high-$n$ DR resonances
\citep[see eg.][]{Savin1997,Savin1999,Schippers1998,Schippers2000b}. Moreover, Rydberg
electrons are more tightly bound in highly charged ions and hence require stronger fields to
become ionized in the charge analyzing field of the apparatus.

The simple picture of a hard cut-off at $n=n_F=19$, as derived in Sec.~\ref{sec:exp}, is not
a good description of the field ionizing properties of our apparatus. This can be seen from
the comparison of the experimental data with the dashed curve in Fig.~\ref{fig:theory} which
has been obtained with $\Upsilon_{nl}=1$ for $n\leq 19$ and $\Upsilon_{nl}=0$ for $n>19$. A
detailed model which takes into account radiative decay of higher Rydberg states on the way
from the cooler to the field ionization magnet is described in appendix \ref{sec:dp}. This
model delivers the $nl$-specific detection probabilities $\Upsilon_{nl}$ plotted in
Fig.~\ref{fig:dp}. Via Eq.~(\ref{eq:faltung}) they yield the dash-dotted curve in
Fig.~\ref{fig:theory}. Apparently, the contributions from Rydberg states with $n>19$ to the
measured rate coefficient are not negligible. Fig.~\ref{fig:dp} indicates that the main
contribution from these high-$n$ Rydberg manifolds is by the short-lived $p$ states. On the
other hand, the cut-off of Rydberg states with $n>45$ in the toroid
(cf.~Tab.~\ref{tab:cutoff}) has a decisive influence on the detection probabilities since
the time available for the radiative decay of higher-$n$ states to below $n=45$ is too short
(see appendix \ref{sec:dp}). The remaining discrepancy between the outcome of the detailed
model and the measured DR spectrum for very high Rydberg states is attributed to model
inherent simplifications. For example, very high-$n$ states are easily perturbed by even
small stray fields possibly leading to enhanced radiative decay rates for $l\neq1$-states
due to $l$-mixing within $n$-manifolds especially in the field-ionization regions. Such
effects are not taken care of by the model.

Finally, we note that at low energies the {\sc autostructure} calculations do not reproduce
the measured $2p4l$ DR resonance structure (inset of Fig.~\ref{fig:theory}), a result which
had to expected given the theoretical effort described by \citet{Mannervik1998} to be
necessary for matching the experiment (inset of Fig.~\ref{fig:overview}).

In view of the substantial experimental cut-off of high Rydberg states we derive our
\ion{C}{4} $\Delta n=0$ DR rate coefficient in a plasma by using the experimental DR
spectrum only below $\hat{E}=7.6$~eV. Above that energy we substitute the experimental DR
spectrum by the scaled {\sc autostructure} result without cut-off (full line in
Fig.~\ref{fig:theory}). The thus created composite DR spectrum is converted into a cross
section $\sigma(\hat{E})=\alpha(\hat{E})/\sqrt{2\hat{E}/m_{\rm e}}$ and convoluted with an
isotropic Maxwellian electron energy distribution yielding the plasma rate coefficient
\begin{mathletters}
\begin{eqnarray}
\alpha(T_{\rm e}) &=& (k_{\rm B}T_{\rm e})^{-3/2}\frac{4}{\sqrt{2 m_{\rm
e}\pi}}\int_0^\infty d\hat{E} \, \sigma(\hat{E}) \hat{E} \exp{(-\hat{E}/k_{\rm B}T_{\rm e})}
\label{eq:alphaTsigma}\\
 & = &  (k_{\rm B}T_{\rm e})^{-3/2}\frac{2}{\sqrt{\pi}}\int_0^\infty
d\hat{E} \, \alpha(\hat{E}) \hat{E}^{1/2} \exp{(-\hat{E}/k_{\rm B}T_{\rm e})}
\label{eq:alphaTalpha}.
\end{eqnarray}
\end{mathletters} at the plasma electron temperature $T_{\rm e}$. This procedure is safe as
long as the relative energy $\hat{E}$ is larger then the experimental energy spread defined
in Eq.~(\ref{eq:deltae}), i.e., for $T_{\rm e} \gg T_{\bot} \approx 120~K$ (see discussion
below).

\begin{figure}
\plotone{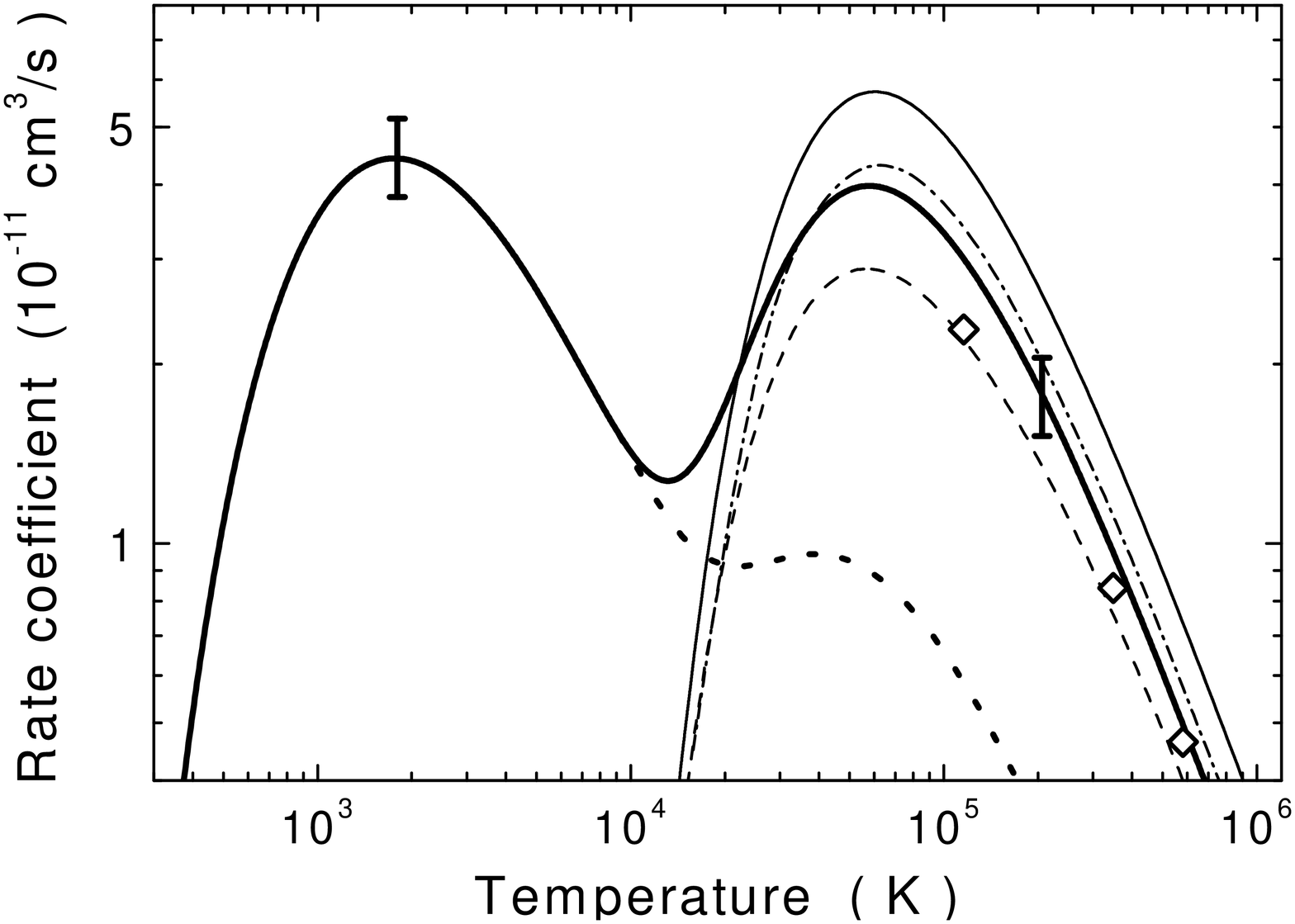} \caption{\label{fig:group1} \ion{C}{4} $\Delta n=0$ DR rate
coefficients in a plasma: this work (thick full line), \protect\citet[][thin full
line]{Burgess1965}, \protect\citet[][dashed line]{Shull1982}, \protect\citet[][dash-dotted
line]{Badnell1989}, and \protect\citet[][open diamonds]{Chen1991}. The dotted curve results
from our measured rate coefficient without the {\sc autostructure} extrapolation. The error
bars represent the $\pm 15\%$ systematic uncertainty of our result.}
\end{figure}

The resulting plasma \ion{C}{4} DR rate coefficient is displayed as the thick full line in
Fig.~\ref{fig:group1}. The curve displays two local maxima. The first one, which is due to
the $2p4l$ DR resonances, is especially benchmarked by our experiment. The second one is
caused by DR via high $n$ Rydberg states. Here our result is dominated by the {\sc
autostructure} calculation that has been adjusted by a constant factor 0.8 to our
experiment. For comparison, the dotted curve in Fig.~\ref{fig:group1} represents the plasma
rate coefficient that has been derived directly from the measured DR spectrum (full symbols
in Fig.~\ref{fig:theory}) without the additional resonance strength above 7.6~eV (shaded
area in Fig.~\ref{fig:theory}) introduced via the {\sc autostructure} calculation . At
temperatures above 10000~K it is about a factor 5 lower than the composite rate coefficient.
On the other hand, the calculated part of the composite DR spectrum does not influence our
result at temperatures below 10000~K.

A convenient representation of the plasma DR rate coefficient is
provided by the following fit formula
\begin{equation}
\alpha(T_{\rm e}) = T_{\rm e}^{-3/2} \sum_i c_i\exp{(-E_i/k_{\rm
B}T_{\rm e})}. \label{eq:alphafit}
\end{equation}
It has the same functional dependence on the plasma electron temperature as the
\citet{Burgess1965} formula, where the coefficients $c_i$ and $E_i$ are related to
oscillator strengths and excitation energies, respectively. The results for the fit to the
experimental \ion{C}{4} $\Delta n=0$ DR rate coefficient in a plasma are summarized in
Table~\ref{tab:fit}. The fit deviates from the thick full line in Fig.~\ref{fig:group1} by
no more than $1\%$ for 650~K$ < T_{\rm e} < $1500~K and by no more than $0.2\%$ for $T_{\rm
e}\geq 1500$~K.

\section{Comparison with theoretical DR rate coefficients}\label{sec:discDR}

\begin{figure}
\plotone{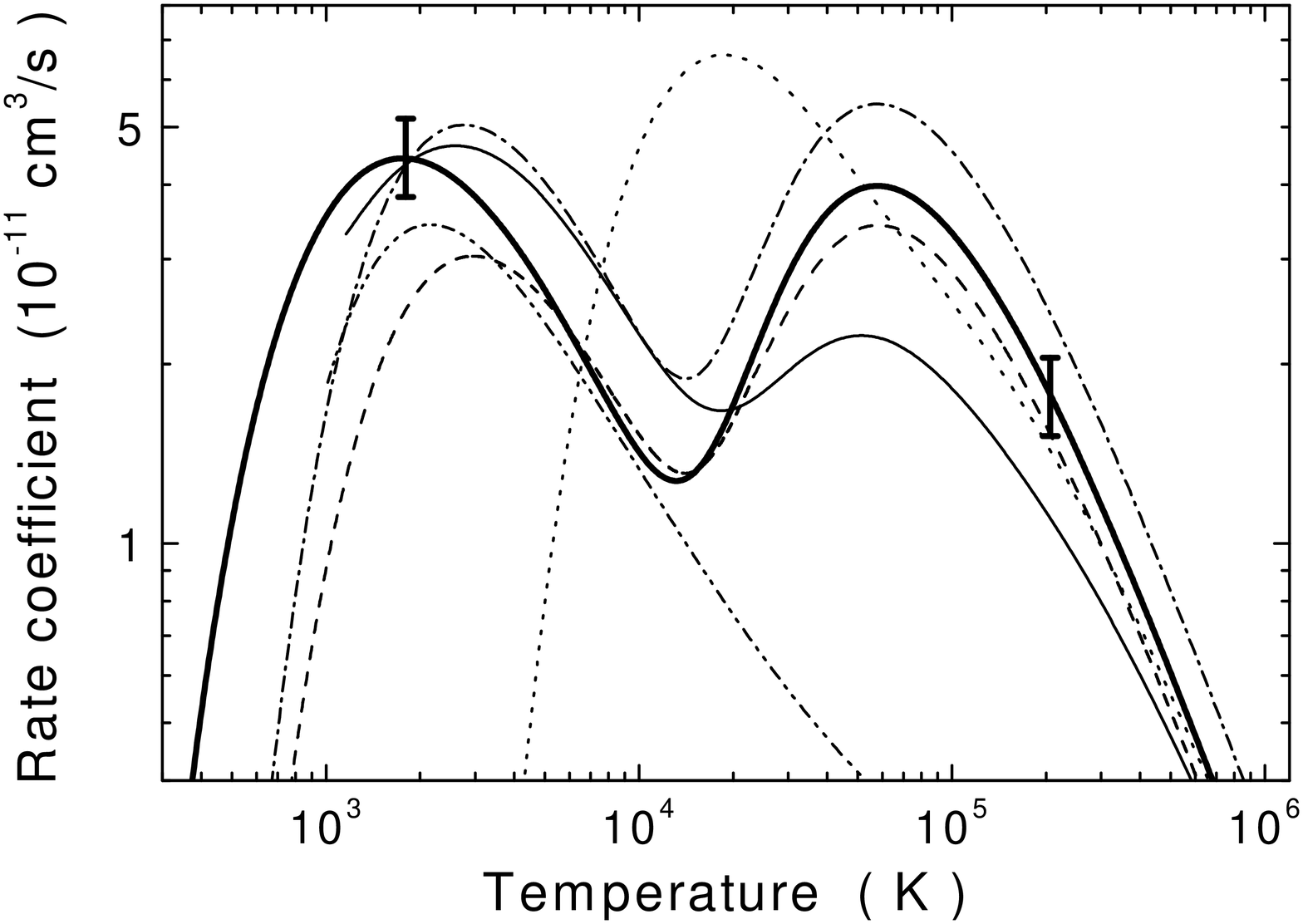} \caption{\label{fig:group2} \ion{C}{4} $\Delta n=0$ DR rate
coefficients in a plasma: this work (thick full line, systematic uncertainty $\pm 15\%$),
\protect\citet[][dashed line]{McLaughlin1983}, \protect\citet[][dash-dot-dotted
line]{Nussbaumer1983}, \protect\citet[][dash-dotted line]{Romanik1988},
\protect\citet[][thin full line]{Safronova1997}, and \protect\citet[][dotted
line]{Mazzotta1998}.}
\end{figure}

Available theoretical \ion{C}{4} DR rate coefficients have been compiled recently by
\citet{Savin2000c}. For a critical assessment of the quality of the various calculations the
reader is referred to that work. Here we confine ourselves to a brief comparison of our
experimentally derived \ion{C}{4} DR plasma rate coefficient with the theoretical results,
which in the following we divide into two categories.

i) The DR calculations by \citet{Burgess1965}, \citet{Shull1982}, \citet{Badnell1989}, and
\citet{Chen1991} have been carried out for high temperatures and do not reproduce the first
local maximum of the experimental rate coefficient below 10000~K (Fig.~\ref{fig:group1}). At
higher temperatures, the \citet{Burgess1965} formula yields rate coefficients up to 50$\%$
larger and at the other extreme the result of \citet{Shull1982} is 30$\%$ lower than the
present one. Above 25000~K the result of \citet{Badnell1989} does not deviate more than
15$\%$ from ours. This deviation is within our experimental uncertainty. The calculation of
\citet{Chen1991} gives rate coefficients only for temperatures higher than $10^5$~K. They
are up to 25$\%$ lower than our rate coefficient in that range.

ii) Calculations that should be valid also at low temperatures have been published by
\citet{McLaughlin1983}, \citet{Nussbaumer1983}, \citet{Romanik1988}, \citet{Safronova1997},
and \citet{Mazzotta1998}. They are shown in Fig.~\ref{fig:group2} together with our result.
At more elevated temperatures above $\sim 20000$~K the rate coefficients of
\citet{Romanik1988} and \citet{Nussbaumer1983} deviate no more from our result than the
group i) calculations discussed in the previous paragraph. In this temperature range the
result of \citet{Safronova1997} is almost a factor 2 lower than ours at temperatures down to
2000~K it is up to a factor 1.5 larger. From 5000 to 5$\times10^5$~K the result of
\citet{McLaughlin1983} is within our error bar of 15$\%$. The rate coefficient of
\citet{Mazzotta1998} agrees reasonably well with our result only above 50000~K. At
temperatures below 2000~K all available calculations deviate strongly from our
experimentally derived rate coefficient. This is most probably due to the neglect of
relativistic and many-body effects which have been shown by \citet{Mannervik1998} to be
essential for the correct description of the DR of even such a light ion as \ion{C}{4}. It
should be mentioned again that at temperatures below $\sim 10^4$~K our result is completely
independent of any theoretical model.

\section{Total recombination rate coefficient}\label{sec:total}

Instead of separate calculations of RR and DR contributions also a unified treatment of both
recombination processes can be considered. Such a treatment, which in principle also
accounts for interference between DR and RR, has been presented by \citet{Nahar1997}.
Experimentally, our measurement also yields the total recombination rate coefficient with
the continuous RR contribution above 0.1~eV accurately represented by the fit described in
section \ref{sec:res}. Below that energy all of the measured rate coefficient
(Fig.~\ref{fig:overview}) is assumed to be exclusively due to RR (see below). In principle,
the total recombination rate coefficient can be derived by re-adding the continuous RR
background to the composite DR spectrum including the {\sc autostructure} calculation based
extrapolation. However, at the low temperatures under consideration two experimental
peculiarities require attention, namely the much discussed \citep{Gao1997,Gwinner2000}
recombination rate enhancement at very low energies and the finite experimental resolution.

\begin{figure}
\plotone{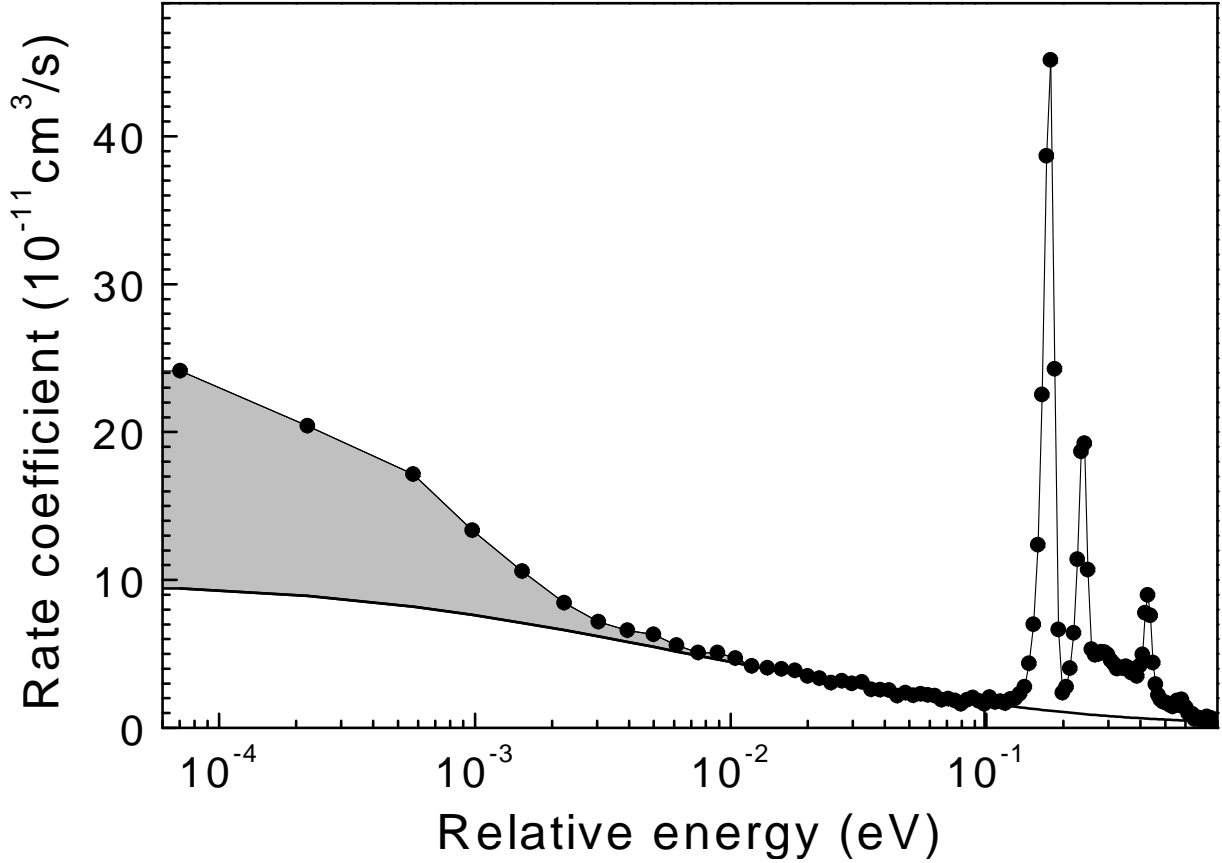} \caption{\label{fig:rr} Measured \ion{C}{4} recombination rate
coefficient at low energies (full circles). The full line is the fitted RR rate coefficient
extrapolated to low energies by a calculation using Eq.~(\protect\ref{eq:sigmaBethe}) (see
text). At energies below 3~meV a strong enhancement of the experimental over the calculated
rate coefficient sets in of up to a factor of about 2.5. The shaded region is the excess
rate coefficient which is neglected in the derivation of the total recombination rate
coefficient in a plasma.}
\end{figure}

The measured enhancement of the recombination rate at very low energies is displayed in
Fig.~\ref{fig:rr}. At energies below 3~meV a strong enhancement of the experimental over the
calculated rate coefficient sets in. At $\hat{E}=0$ this factor reaches a value of about
2.5. This effect has been found in electron-ion recombination measurements at different
storage rings and has not been explained yet. Systematic studies of this effect have been
carried out by \citet{Gao1997} and \citet{Gwinner2000}. The recombination rate enhancement
results most probably from the specific experimental arrangement at storage-ring electron
coolers where the electron beam is guided by a magnetic field.

Since we do not expected the enhancement to occur in an astrophysical environment we
subtract the excess rate coefficient (shaded area in Fig.~\ref{fig:rr}) from our
experimental data. To this end we have extrapolated the fitted RR background (see
Sec.~\ref{sec:res}) to lower energies by scaling a RR rate coefficient that has been
calculated using a modified version of the semi-classical \citet{Bethe1957} formula for the
hydrogenic RR cross section, i.e.
\begin{equation}\label{eq:sigmarr}
\sigma^{RR}(\hat{E}) = 2.10\times10^{-22}{\rm cm}^2\times \sum_{n=n_{\rm min}}^{n_{\rm max}}
t_n G_n(0) \frac{q^4 {\cal R}^2}{n \hat{E}\, (q^2 {\cal R} + n^2
\hat{E})}\label{eq:sigmaBethe}
\end{equation}
with the Rydberg constant ${\cal R}$. For RR onto \ion{C}{4} $q=3$ and $n_{\rm min}=2$ are
appropriate. As discussed in Sec.~\ref{sec:exp} the maximum quantum number to be taken into
account is determined by field ionization in our experimental setup. Since a rough estimate
is sufficient for the present purpose we take $n_{\rm max} =20$. The factors $t_n$ account
for partially filled shells. Here we use $t_2 = 7/8$ and $t_n = 1$ for $n\geq 3$. The Gaunt
factors $G_n(\hat{E})$ are small corrections which account for deviations of the
semi-classical formula from the quantum mechanically correct hydrogenic result. Generally
the Gaunt factors are weakly energy dependent. We have taken energy independent values
calculated for $\hat{E}=0$ by \citet{Andersen1990}.

\begin{figure}
\plotone{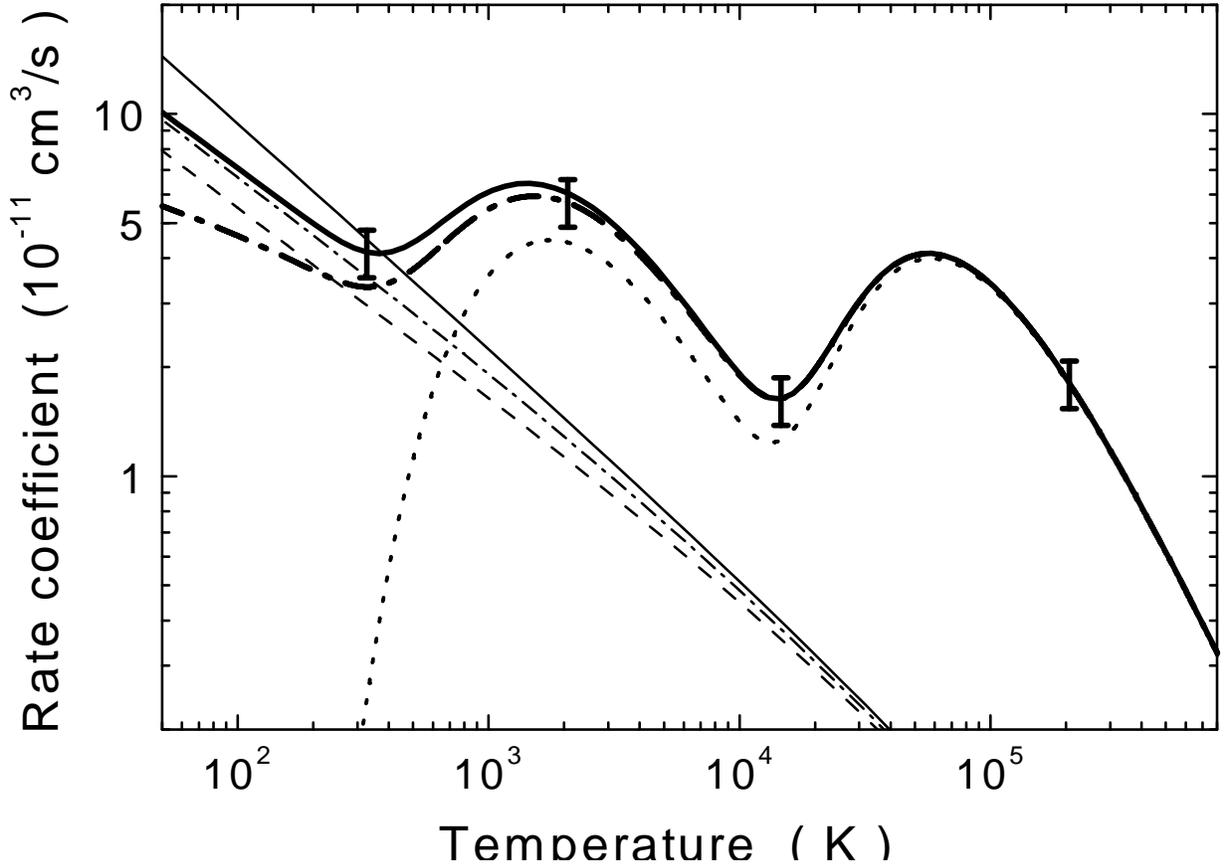} \caption{\label{fig:unified} Experimental total \ion{C}{4}
recombination rate coefficients in a plasma (thick full line, systematic error $\pm 15\%$).
The comparison with our pure DR rate coefficient (dotted line) shows that RR is noticeable
up to $\sim 30000$~K. The thick dash-dotted line is our total recombination rate coefficient
uncorrected for the influence of the finite experimental resolution. The other lines are
\ion{C}{4} RR rate coefficients of \protect\citet[][thin full line]{Pequignot1991} and
corresponding RR rate coefficients (see text) for $n_{\rm max}= 20$ (thin dashed line) and
$n_{\rm max}= 40$ (thin dash-dotted line).}
\end{figure}

The thus calculated RR cross section has been convoluted by the experimental electron
velocity distribution [see Eqs.~(\ref{eq:fele}) and (\ref{eq:faltung})] and the resulting
rate coefficient has been multiplied by a constant factor of 1.638 such that it matches the
fitted RR background at $\hat{E}= 0.1$~eV. The finding that a factor different from unity
has to be used in order to achieve the matching, can be attributed to the fact that the
hydrogenic treatment is not appropriate for RR at least into low-$n$ shells of the
lithiumlike \ion{C}{4} ion. The total recombination rate coefficient is now calculated as
the sum of the extrapolated RR background (full line in Fig.~\ref{fig:rr}) and the DR rate
coefficient derived in Sec.~\ref{sec:res} (cf.\ Tab.~\ref{tab:fit}). This ensures that the
excess rate due to the recombination rate enhancement does not contribute to the total
recombination rate coefficient which is displayed as the thick dash-dotted line in
Fig.~\ref{fig:unified}.

At low plasma temperatures we also have to consider the influence of the finite experimental
resolution [cf.~Eq.~(\ref{eq:deltae})] on our total plasma rate coefficient, which is only
well defined at temperatures $T_{\rm e} \gg 120$~K as discussed before (Sec.\ref{sec:res}).
In order to quantify this influence we have convoluted the theoretical RR cross section
[Eq.~(\ref{eq:sigmaBethe}) with $n_{\max}=20$] with our experimental energy distribution
[see Eqs.~(\ref{eq:fele}) and (\ref{eq:faltung})] and derived from the resulting RR rate
coefficient $\alpha^{RR}(\hat{E})$ via Eq.~(\ref{eq:alphaTalpha}) the "doubly convoluted"
plasma RR rate coefficient. This is to be compared with the "singly convoluted" RR rate
coefficient
\begin{equation}
 \alpha^{RR}(T_{\rm e})
 = 5.20\times10^{-14}\mbox{\rm cm}^3\mbox{\rm s}^{-1} \times
 q\,\sum_{n=n_{\rm min}}^{n_{\rm max}}t_n G_n(0)\;\Theta_n^{3/2}\;\exp(\Theta_n)\,{\rm E_1}(\Theta_n) \label{eq:RRplasma}
\end{equation}
where $E_1(\Theta_n) = \int_{\Theta_n}^\infty x^{-1}\exp(-x){\rm d}x$ is the exponential
integral and $\Theta_n = q^2{\cal R}/(n^2 k_{\rm B}T_{\rm e})$. Eq.~(\ref{eq:RRplasma}) has
been derived by inserting Eq.~(\ref{eq:sigmaBethe}) into Eq.~(\ref{eq:alphaTsigma})
\citep[see also][]{Seaton1959}. In contrast to the "doubly convoluted" rate coefficient the
"singly convoluted" one is not influenced by the experimental resolution. We find that at
$T_{\rm e} = 500$~K (200~K) the "doubly convoluted" rate coefficient is $19\%$ ($35\%$)
lower than the "singly convoluted" one. It should be noted that these numbers depend
strongly on the shapes of the convoluted curves. Flatter curves than the diverging RR cross
sections will yield much smaller values. In order to correct for the influence of the finite
experimental resolution we have multiplied our previously derived total \ion{C}{4}
recombination rate coefficient (dash-dotted curve in Fig.~\ref{fig:unified}) with the
temperature dependent ratio of the "singly" to "doubly convoluted" RR rate coefficient. The
resulting corrected total recombination rate coefficient is displayed as the thick full line
in Fig.~\ref{fig:unified}. In principle the correction also influences the DR rate
coefficient. However, in the relevant temperature range the correction is well within the
overall systematic error and its influence on the \ion{C}{4} plasma DR rate coefficient has
therefore been neglected.

At low temperatures our experimental total rate coefficient stays below the widely used
theoretical RR rate coefficient given by \citet{Pequignot1991} (thin full line in
Fig.~\ref{fig:unified}). This is due to the field ionization of high Rydberg states. In
their calculation \citet{Pequignot1991} account for RR into all $n$-shells up to infinite
$n$. For $ 4\leq n < \infty$ they use the hydrogenic RR rate coefficient of
\citep{Martin1988}. In order to compare our result to theoretical rate coefficients for
finite values of $n_{\rm max}$ we first subtract this hydrogenic contribution from the rate
coefficient of \citet{Pequignot1991} thereby retaining their non-hydrogenic \ion{C}{4} RR
rate coefficient with $n_{\rm max}=3$. Onto this we have then added the hydrogenic RR rate
coefficient calculated by using Eq.~(\ref{eq:RRplasma}) with $n_{\rm min}=4$, $n_{\rm
max}=20$ and $n_{\rm min}=4$, $n_{\rm max}=40$ yielding the thin dashed and dash-dotted
curves in Fig.~\ref{fig:unified}, respectively. Extending the summation to $n_{\rm
max}=1000$ yields --- as expected --- a curve which is indistinguishable from the result of
\citet{Pequignot1991} on the scale of Fig.~\ref{fig:unified}. At low temperatures our
experimental rate coefficient approaches the theoretical curve that has been obtained with
$n_{\rm max} = 40$. This is roughly consistent with the prediction of the detailed model for
field ionization in our experimental apparatus as discussed in Sec.~\ref{sec:exp} and in the
appendix. From this observation we conclude that within our 15~\% systematic uncertainty the
total \ion{C}{4} recombination rate coefficient can be represented as the sum of the RR rate
coefficient of \citet{Pequignot1991} and our DR rate coefficient parameterized by
Eq.~(\ref{eq:alphafit}) with the parameters listed in Tab.~\ref{tab:fit}.

\begin{figure}
\plotone{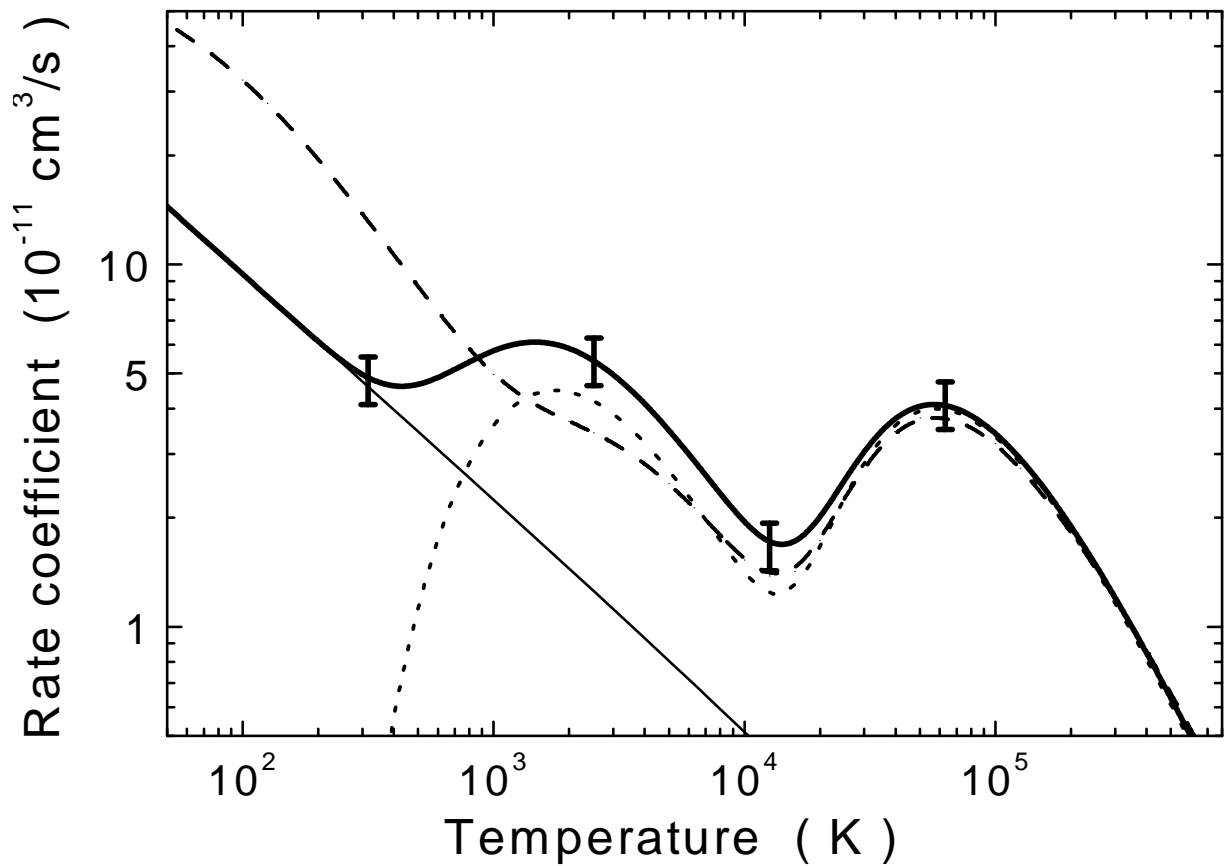} \caption{\label{fig:total} Total \ion{C}{4} recombination rate
coefficients in a plasma: this work (thick full line, systematic error $\pm 15\%$) and
theoretical unified calculation of \citet{Nahar1997} (dashed line). Our total recombination
rate coefficient is obtained as the sum of the RR rate coefficient of \citet[][thin full
line]{Pequignot1991} and our DR rate coefficient (dotted line) [see
Eq.~(\protect\ref{eq:alphafit}) and Tab.~\protect\ref{tab:fit}].}
\end{figure}

In Fig.~\ref{fig:total} this total \ion{C}{4} recombination rate coefficient is compared to
the results of the unified RR+DR calculation of \citet{Nahar1997}. At temperatures above
5000~K both results agree with each other within our systematic uncertainty. At lower
temperatures, however, pronounced differences occur. The theoretical result is up to a
factor of 3 larger than the experimental one. It should be noted that interference between
RR and DR cannot be held responsible for the observed discrepancy, since it is included both
in the unified theory and in the experiment. The measured rate coefficient especially fully
includes the strong $2p4l$ DR channels. Only these can in principle be expected to give rise
to noticeable interference effects at low plasma temperatures. However the lowest
experimental \ion{C}{4} DR resonance occurs at about $0.18$~eV. Assuming a delta-like
cross-section and using the (somewhat to large) theory value of $\sim
5\times10^{-19}$~cm$^2$eV for its strength \citep{Mannervik1998} we calculate its
contribution to the total rate coefficient by the use of Eq.~(\ref{eq:alphaTsigma}). At
500~K it is $\sim 1\times10^{-12}$~cm$^3$/s, i.e., about $2\%$ only. Accordingly, at lower
temperatures the measured rate coefficient should be almost entirely due to RR, i.e., the
low-temperature deviation of the unified theory of \citet{Nahar1997} from our measured
\ion{C}{4} rate coefficient remains to be explained.

\section{Summary and conclusions}

We have measured the $\Delta n=0$ \ion{C}{4} DR rate coefficient by detecting with
essentially full geometric efficiency the recombination products following interaction in
merged electron and ion beams. Generally, in such experiments the field ionization related
to the charge analysis of the recombination products causes a large fraction of the DR
resonance strength due to DR via high Rydberg states to remain unmeasured. For the present
case of $\Delta n=0$ DR of a low-charge ion this undetected DR strength is substantial. In
order to provide a remedy for this deficiency, we have performed {\sc autostructure}
calculations, carefully modeled the Rydberg-ion detection probabilities of our apparatus,
scaled the results to our experimental low energy DR rate coefficient, and finally used the
high-energy part of the calculation as a substitution for the unmeasured DR strength. From
the thus extrapolated experimental DR spectrum we have derived the \ion{C}{4} DR rate
coefficient in a plasma [cf.\ Eq.~(\protect\ref{eq:alphafit}) and
Tab.~\protect\ref{tab:fit}]. Furthermore the careful analysis of our experimental data leads
to the conclusion that within our systematic uncertainty of 15 \% the total \ion{C}{4}
recombination rate coefficient can be represented as the sum of the DR rate coefficient and
the theoretical RR rate coefficient of \citet{Pequignot1991}. This implies that interference
between DR and RR is insignificant for the recombination of \ion{C}{4}. We have compared our
rate coefficients with the available theoretical results. None of them agrees with our
experimental rate coefficients over the full temperature range.

The field ionization of recombined ions in high Rydberg states, which is unavoidable in ion
storage-ring experiments, ultimately limits the capability of providing meaningful, pure
experimental plasma DR rate coefficients. This is especially true for low charge-state ions
where much DR strength is concentrated in high-$n$ resonances. Due to the much higher
radiative rates \citep[$Z^4$-scaling,][]{Bethe1957} the DR resonance strength drops much
faster with increasing $n$ for high charge-state ions so that a smaller fraction of the
total DR strength is accumulated in high Rydberg states. Consequently, storage-ring
experiments can provide reliable DR plasma rate coefficients with much more limited need of
extrapolation for highly charged ions.

\acknowledgments

Technical support by the Heidelberg accelerator group and the TSR team is gratefully
acknowledged. We thank D.~W.\ Savin for stimulating discussions and for providing to us his
compilation of theoretical \ion{C}{4} DR rate coefficients in numerical form. We also thank
N.~R.\ Badnell for making the {\sc autostructure} code available via the world wide web.

\appendix

\section{Model calculation of detection probabilities for high Rydberg states}
\label{sec:dp}

\begin{figure}
\plotone{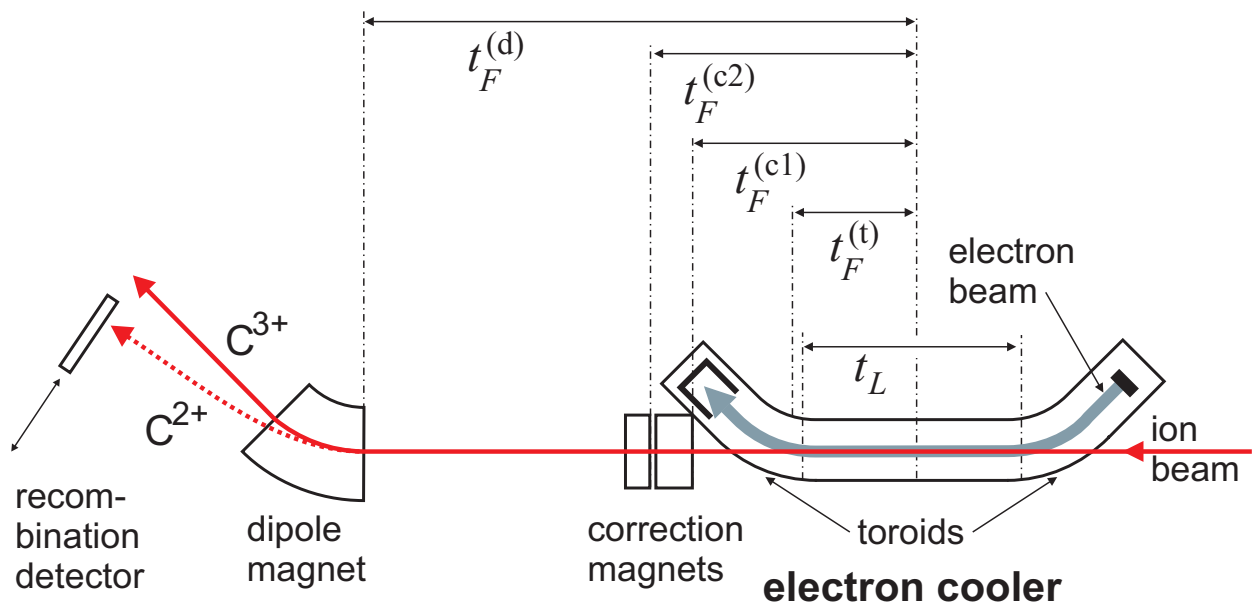} \caption{\label{fig:exp}Sketch of the experimental setup. The C$^{3+}$
ion beam enters the electron cooler from the right. Behind the merging section inside the
cooler the ions pass the toroidal magnet used for deflecting the electron beam out of the
ion beam's path, two correction dipole magnets and the charge analyzing dipole magnet. The
respective flight times, $t_F^{\rm (t)}=55$~ns, $t_F^{\rm (c1)}=103$~ns, $t_F^{\rm
(c2)}=133$~ns, and $t_F^{\rm (d)}=275$~ns, from the center of the electron cooler to these
magnets are indicated; $t_L=87$~ns is the flight time through the merging section.
Recombined C$^{2+}$ ions are counted with the recombination detector. Not shown are
correction magnets to the right of the cooler and focusing magnets in between the correction
and charge analyzing dipoles.}
\end{figure}

In the detailed model of the field ionization properties of our
apparatus, the survival fractions are determined individually for
each $nl$ state populated by DR, in combination with the flight
times $t_F$ to the field ionization zones and hydrogenic decay
rates. Assuming a constant recombination probability across the
length $L$ of the merging section the probability that a state
characterized by quantum numbers $n$ and $l$ has decayed upon
reaching the field ionization zone is given as
\begin{equation}
P_{\rm d}(nl,t_L,t_F) =
1-\frac{\tau(nl)}{t_L}\left[\exp\left(-\frac{t_F-t_L/2}{\tau(nl)}\right)-
\exp\left(-\frac{t_F+t_L/2}{\tau(nl)}\right)\right].\label{eq:pd}
\end{equation}
The flight time through the electron cooler is $t_{L}=87$~ns in the present case and the
radiative lifetime of the $nl$-state is calculated as $\tau(nl) =
[\sum_{n'<n,l'=l\pm1}\gamma_{\rm r}(nl\to n'l')]^{-1}$ from hydrogenic radiative dipole
transition rates $\gamma_{\rm r}(nl\to n'l')$ following \citet{Bethe1957}. In
Eq.~(\ref{eq:pd}), which is valid for $t_F \geq t_L/2$, $t_F$ is measured from the center of
the merging section inside the electron cooler up to the field ionization region
(cf.~Fig.~\ref{fig:exp}). There the recombined ion spends a time $\Delta t_F$ during which
field ionization may take place. The ion's survival probability is calculated from
approximate field ionization rates $A_F$ derived by \citet{Damburg1979} to be
\begin{equation}
 P_{\rm s}(nl,\Delta t_F,F)=\frac{1}{2l+1}\sum_{m=-l}^l \sum_{n_1=0}^{n-\vert m\vert-1}
 \left(C^{n,l}_{n_1,m}\right)^2
 \exp[-\Delta t_F A_F(n_1,n_2,m)]\label{eq:ps}
\end{equation}
where the expansion coefficients are the Clebsch-Gordan
coefficients
\begin{equation}
 C^{n,l}_{n_1,m} = \Biggr\langle\begin{array}{cc} (n\!-\!1)/2 &  (n\!-\!1)/2\\
 (m+\!n_1\!-\!n_2)/2 & (m\!-\!n_1\!+\!n_2)/2\end{array}\Biggr\vert
 \Biggl.\begin{array}{c}l\\m\end{array}\Biggr\rangle
\end{equation}
for the transformation to Stark states characterized by the
parabolic quantum numbers $n_1$, $n_2$ and $m$, which satisfy the
relation $n = n_1+n_2+\vert m\vert +1$ \citep{Gallagher1994}.
Finally, the detection probability is calculated from the decay
and survival probabilities as
\begin{eqnarray}
\Upsilon_{nl}(t_L,t_F,\Delta t_F,F) &=& [1-P_{\rm d}(nl,t_L,t_F)]\,P_{\rm s}(nl,\Delta
t_F,F)+\sum_{n'<n,l'=l\pm 1}b(nl\to n'l')\times\nonumber\\&& \biggl\{\,[P_{\rm
d}(nl,t_L,t_F)\!-\!P_{\rm d}(nl,n'l',t_L,t_F)]\,P_{\rm s}(n'l',\Delta t_F,F)\nonumber\\ && +
 \,P_{\rm c}(nl,n'l',\dots,t_L,t_F,\Delta t_F,F)\biggr\}
 \label{eq:efficiency}
\end{eqnarray}
where branching ratios for a dipole transition from state $nl$ to
state $n'l'$ are defined as $b(nl \to n'l') =
\tau(nl)\,\gamma_{\rm r}(nl\to n'l')$. The quantity
\begin{eqnarray}
P_{\rm c}(nl,n'l',&\ldots&,t_L,t_F,\Delta t_F,F) = \sum_{n''<n',l''=l'\pm 1}\!\!\! b(n'l'\to
n''l'')\label{eq:pc}\\ &\times\,\biggl\{&[P_{\rm d}(nl,n'l',t_L,t_F)\!-\!P_{\rm
d}(nl,n'l',n''l'',t_L,t_F)]\nonumber\\&&\times\,P_{\rm s}(n''l'',\Delta t_F,F) +P_{\rm
c}(nl,n'l',n''l'',\dots,t_L,t_F,\Delta t_F,F)\biggr\}\nonumber
\end{eqnarray}
accounts for cascading, i.e.\ stepwise de-excitation by more than
one transition. It is calculated recursively. In
Eqs.~(\ref{eq:efficiency}) and (\ref{eq:pc})
\begin{equation}
P_{\rm d}(n_1l_1,\ldots,n_Nl_N,t_L,t_F) = \sum_{k=1}^N
\frac{[\tau(n_kl_k)]^{N-1}}
                 {\prod_{j \neq k}[\tau(n_k,l_k)-\tau(n_j,l_j)]}\;P_{\rm d}(n_kl_k,t_L,t_F)
                 \label{eq:pdc}
\end{equation}
is the probability for a decay along a given sequence $n_1l_1
\dots n_Nl_N$ of $N$ excited hydrogenic states
\citep{Schippers1995}. The one-step decay probabilities $P_{\rm
d}(n_kl_k,t_L,t_F)$ are evaluated according to Eq.~(\ref{eq:pd}).

\begin{figure}
\plotone{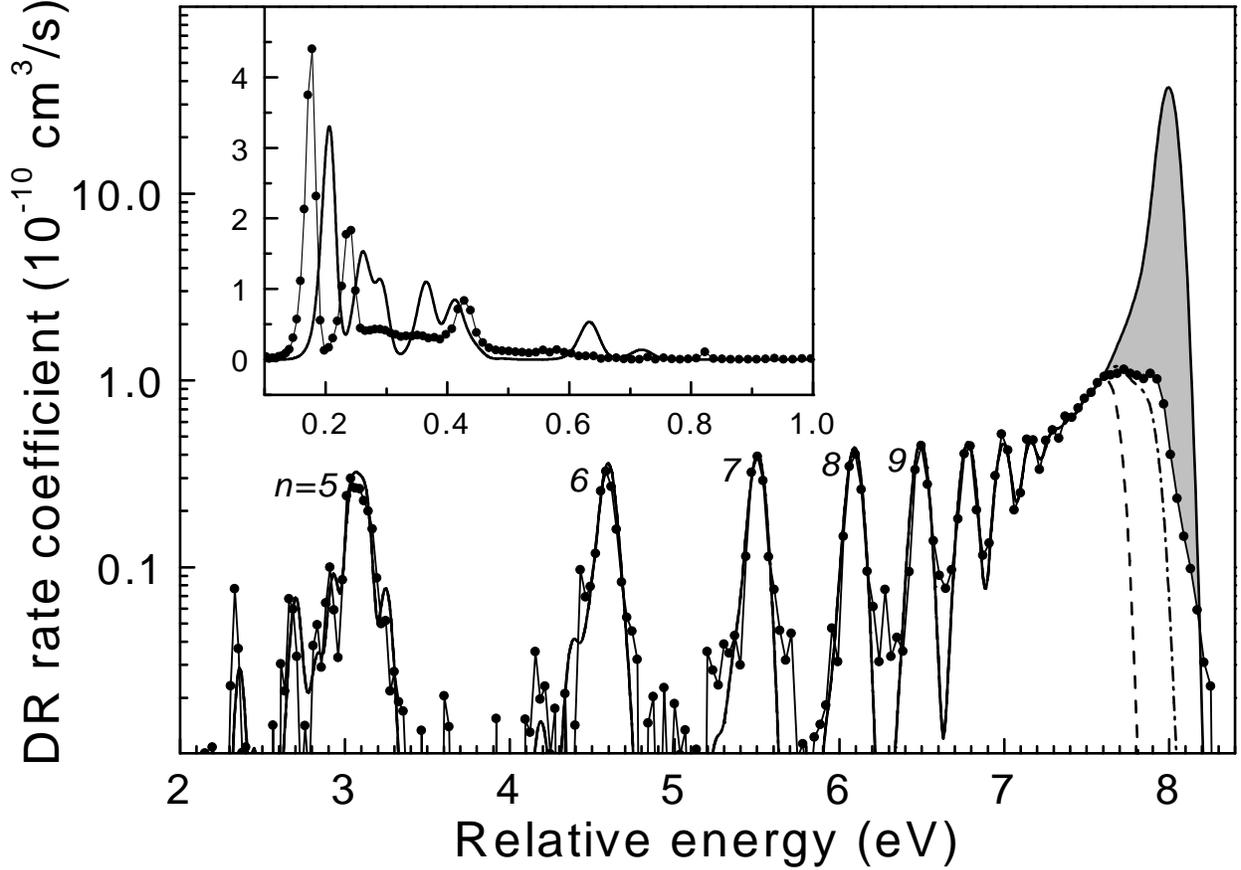} \caption{\label{fig:theory} Comparison between experiment (full
circles) and our {\sc autostructure} calculation. The calculated rate coefficient has been
multiplied by a factor 0.8 and the theoretical energy scale has been shifted by 0.06~eV
towards higher energies (see text). The shaded area highlights the unmeasured, purely
calculated part of the composite DR rate coefficient. The different curves correspond to
different assumptions for the cut-off of high Rydberg states, i.e., no cut-off (full line),
cut-off at $n = n_F = 19$ (dashed line) and detailed model (dash-dotted line, see text). The
inset shows the comparison in the region of the $2p4l$ DR resonances.}
\end{figure}

Due to the high $n$ values to be considered (up to $n \approx 100$), the explicit
calculation of cascade contributions is very laborious and time consuming. In view of the
fact that rates are highest for transitions to the lowest available state and decrease with
$\sim n'^{-3}$ \citep{Bethe1957} it can be anticipated that except for high angular momentum
states, which are only sparsely populated by DR, cascading only plays a minor role. We have
verified this by comparing calculations without and with cascades via one and two
intermediate states. The resulting rate coefficients are undistinguishable on the scale of
Fig.~\ref{fig:theory}. Therefore, we conclude that it is safe to neglect contributions from
cascades with more intermediate states.

\begin{figure}
\plotone{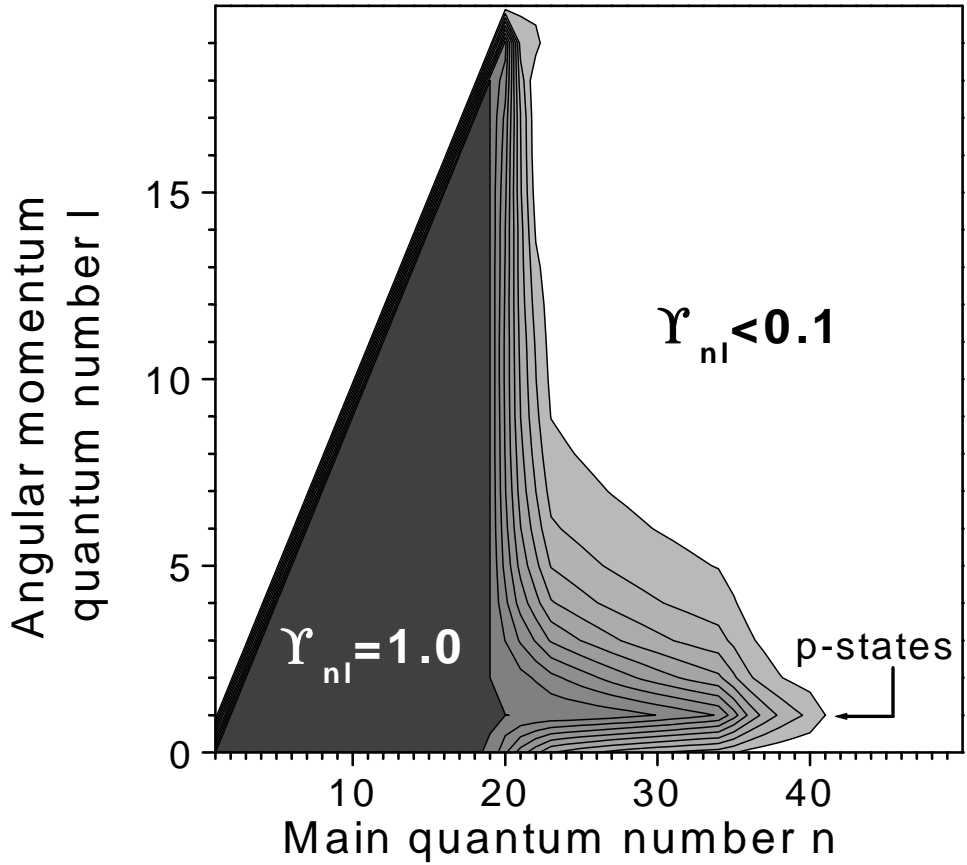} \caption{\label{fig:dp} Contour plot of
detection probabilities from the detailed model. The contours are
(from right to left) $\Upsilon_{nl}=0.1, 0.2, 0.3\ldots 1.0$. A
considerable fraction of low $l$ Rydberg states is detected with
high probability even for high $n$. The region $l \geq n$ is
unphysical.}
\end{figure}

In the calculation of detection probabilities we include states up to $n=100$ which is far
beyond the estimated hard cut-off at $n_F=19$ (cf.~Sec.~\ref{sec:exp}). Furthermore, we
consider all field ionization regions which have been mentioned previously: the toroidal
section at the exit of the electron cooler, two correction magnets and the charge analyzing
dipole magnet (cf.\ Fig.~\ref{fig:exp}). The corresponding detection probabilities
$\Upsilon^{\rm (t)}_{nl}$, $\Upsilon^{\rm (c1)}_{nl}$, $\Upsilon^{\rm (c2)}_{nl}$ and
$\Upsilon^{\rm (d)}_{nl}$, respectively, are calculated individually and the overall
detection probability is finally given as the product of the four individual ones, i.e.
\begin{eqnarray}
 \Upsilon_{nl} &=&
 \Upsilon_{nl}^{\rm (t)}\left(t_L,t_F^{\rm (t)},\Delta t_F^{\rm (t)},F^{\rm(t)}\right)\times
 \Upsilon_{nl}^{\rm (c1)}\left(0,t_F^{\rm (c1)}-t_F^{\rm (t)},\Delta t_F^{\rm (c1)},F^{\rm (c1)}\right)\nonumber\\
 &&\times\, \Upsilon_{nl}^{\rm (c2)}\left(0,t_F^{\rm (c2)}-t_F^{\rm (c1)},\Delta t_F^{\rm (c2)},F^{\rm (c2)}\right)
\times \Upsilon_{nl}^{\rm (d)}\left(0,t_F^{\rm (d)}-t_F^{\rm
(c2)},\Delta t_F^{\rm (d)},F^{\rm (d)}\right)
 \label{eq:eff4}
\end{eqnarray}
with the numerical values for $t_F$, $\Delta t_f$ and $F$ taken from Tab.~\ref{tab:cutoff}.
It should be noted, that for the calculation of $\Upsilon_{nl}^{\rm (c1)}$,
$\Upsilon_{nl}^{\rm (c2)}$ and $\Upsilon_{nl}^{\rm (d)}$ we used the $t_L \to 0$ limit of
Eq.~(\ref{eq:pd}), namely $P_{\rm d}(nl,0,t_F) = 1-\exp[-t_F/\tau(nl)]$. A contour plot of
the calculated detection probabilities is shown in Fig.~\ref{fig:dp}. The detection
probabilities are 100$\%$ for $nl$-states up to $n=19$. At higher $n$ only low $l$ states
have a high detection probability owing to their short lifetimes. Higher $l$ states have
increasingly higher lifetimes and are more and more effectively field ionized. Consequently,
for $n>19$ the detection probability drops very fast to zero with increasing $l$. Above
$n\approx 40$ also low $l$ states are cut-off due to field ionization in the toroid.
According to Eq.~(\ref{eq:nf}) the hard cut-off number there is $n_F=45$. Due to the short
time of only 55~ns needed for travelling from the cooler center to the toroid almost no
radiative de-excitation of higher Rydberg states is possible.


\clearpage

\begin{deluxetable}{ccc}

\tablecolumns{5} \tablewidth{0pt} \tablecaption{Fit parameters for the experimentally
inferred \ion{C}{4} $\Delta n=0$ DR rate coefficient.\label{tab:fit}}

\tablehead{\colhead{$i$} & \colhead{$c_i$} & \colhead{$E_i$}}

 \startdata
 1 & 2.420E-3 & 7.969 \\
 2 & 1.347E-4 & 5.386 \\
 3 & 1.094E-5 & 0.330 \\
 4 & 6.897E-6 & 1.650 \\
 5 & 6.328E-6 & 0.169 \\
 \enddata
\tablecomments{Units are cm$^3$s$^{-1}$K$^{1.5}$ for $c_i$, and eV for $E_i$. The systematic
error of the rate coefficient $\alpha(T_{\rm e})$ from Eq.~(\protect\ref{eq:alphafit}) is
$\pm 15\%$.}
\end{deluxetable}

\begin{deluxetable}{lcccc}

\tablecolumns{5} \tablewidth{0pt} \tablecaption{Field ionization
regions in the flight path of recombined ions\label{tab:cutoff}}

\tablehead{\colhead{Region}& \colhead{$F$} & \colhead{$\Delta
t_F$} & \colhead{$t_F$} & \colhead{$n_F$}}
 \startdata
 toroid                & \phn\phn  4    &  29          & \phn 55 & 45    \\
 1st correction magnet & \phn\phn  6    &  20          &   103   & 40    \\
 2nd correction magnet &     \phn 12    &  10          &   133   & 34    \\
 dipole magnet         &         106    &  49          &   275   & 19    \\
 \enddata
\tablecomments{Units are kV/cm for $F$ and ns for $\Delta t_F$ and
$t_F$. $n_F$ has been calculated from Eq.~(\protect\ref{eq:nf}).}
\end{deluxetable}

\clearpage


\end{document}